\documentclass[12pt]{article}
\usepackage{graphicx}
\def\Red  {}
\def\Black{}
\def\Blue {}

\begin{document}

\centerline{ hep-ph/9912504 \hfill Published in Phys. Lett. {\bf B477}, 223 (2000)}


\renewcommand{\thefootnote}{\fnsymbol{footnote}}
\setcounter{footnote}{-1}

\vspace*{15mm}

\begin{center}\Red
{\Large\bf Higgs Mass Prediction with Non-universal \\[0.3cm]
       Soft Supersymmetry Breaking in MSSM} \\[1cm]%
\Black
{\bf
Sorin~Codoban\footnote[1]{\makebox[1.cm]{E-mail:}
                codoban@thsun1.jinr.ru},
Marian~Jur\v{c}i\v{s}in
\footnote[2]{\makebox[1.cm]{E-mail:}
jurcisin@thsun1.jinr.ru}$^,$\footnote[3]{On leave of absence from
 the Institute of
 Experimental Physics SAS,  Ko\v{s}ice, Slovakia}
and Dmitri~Kazakov
\footnote[4]{\makebox[1.cm]{E-mail:}
kazakovd@thsun1.jinr.ru}} \\ [5mm]

{\em Bogoliubov Laboratory of Theoretical Physics,
JINR , 141 980 Dubna, Russia}%
\end{center}
\vspace*{0.5cm}



\begin{abstract}\Blue
In the framework  of the MSSM the non-universal boundary
conditions of soft SUSY breaking parameters are considered.
Taking as input the top, bottom and Z-boson masses,
the values of the gauge couplings at the EW scale and the
infrared quasi-fixed points for Yukawa couplings and the
soft parameters the mass of the lightest CP-even Higgs
 boson is found to be
$m_h=92.7^{\displaystyle +10}_{\displaystyle -4.9} \pm 5\,\pm 0.4\
GeV/c^2$ for the low $\tan\beta$ case and $m_h
=125.7^{\displaystyle +6.4}_{\displaystyle -9.0}\pm 5\, \pm 0.4 \
GeV/c^2$ ($\mu > 0)$ or  $m_h =125.4^{\displaystyle
+6.6}_{\displaystyle -9.0}\pm 5\, \pm 0.4 \ GeV/c^2$ ($\mu < 0)$
 in the case of large  $\tan\beta$.
\end{abstract}

\Black
\section{Introduction}
 When making predictions in the framework of the Minimal
Supersymmetric Extension of the Standard Model (MSSM), one encounters
parameter freedom which is mainly due to the so-called soft SUSY
breaking terms.
 To restrict this freedom and  get more
predictive power, one usually follows the universality hypothesis
that assumes  universality (i.e. equality)  of  soft terms at
some high energy scale (most common is the GUT scale at which the
gauge couplings are unified).
 Under this assumption one is left with 5 free parameters,
and a thorough analysis of the MSSM mass spectrum in universal case
has  been done \cite{uniMSSM}.

 However, the newest experiments, and first of all recent LEP II
limits on the lightest Higgs boson mass \cite{LEP_new_results},
suggest that this minimal scenario might not work in practice.
 In the present paper we try to clarify
what one  expects to get by releasing the universality
conditions in the MSSM, allowing more freedom for the soft terms,
with the emphasis on the  Higgs boson mass predictions.

 In order to see how the non-universality at the GUT scale can
change the predictions at the low energy scale, we use  recently
obtained analytical solutions to the renormalization group (RG)
equations for the Yukawa couplings \cite{moultaka,KM}.
 By means of Grassmannian expansion they allow one to derive
analytical expressions for soft term evolution
 \cite{kaz_avd_kond,kaz_physlettB} and trace analytically the
dependence of the soft terms at  the $M_Z$ scale on their boundary
values.
 In this way, both the expediency of one or another
simplifying hypotheses concerning GUT scale values of the soft
parameters and the role of non-universality in the context of the
MSSM  can be estimated.

 The parameter space of the MSSM can further be narrowed using the
so-called infrared quasi-fixed point (IRQFP) behaviour of some
parameters \cite{Hill}, i.e. independence of low energy values on
initial conditions at the GUT scale.
 Using the analytical results along with
the numerical ones, one can keep over control the way how the
IRQFP strategy works. In what follows we adopt the method
advocated in Refs. \cite{YJK,JK} and apply the above-mentioned tips
in making prediction for the lightest Higgs boson mass in the MSSM.

 The paper  is organized as follows.
In Sec.~\ref{ir_analytic}, we analyze the IRQFP behaviour of soft
parameters with the help of analytical solutions.
 In Sec.~\ref{lowtan}, the low $\tan\beta$ scenario with
non-universality is investigated, and prediction for Higgs boson
mass  is given.
 The same analysis is conducted in
Sec.~\ref{bigtan} for the case of large $\tan\beta$.
Our concluding remarks are presented in Sec.~\ref{the_end}.

\section{
Analytical solutions of RG equations  for the soft terms and IRQFPs
}
\label{ir_analytic}

 As has been recently shown \cite{moultaka}, the RGEs for Yukawa
couplings can be solved analytically  by means of an iteration
procedure.
 Following the recipe advocated in \cite{KM,kaz_physlettB},
the expressions for soft parameters can be
derived from analytical solutions of Yukawa coupling RGEs.
 Below we provide a brief analysis of analytical solutions for the
soft parameters, emphasizing the difference between universal and
non-universal cases.
 For low $\tan\beta$ an exact analytical solution is known
\cite{exactlow,bottom-up,kaz_cod}; so we discuss  just the case of
large $\tan\beta$ when all the Yukawa couplings  are essential.

  Consider solutions to the RG equations for the soft terms.
 Since the triple scalar couplings $A_{t,b,\tau}$ and gaugino masses
$M_i$ have a dimension of a mass and the squark, slepton and Higgs
mass terms have a dimension of a mass squared,  the corresponding
RG equations are linear with respect to these parameters, and
their solutions can be represented in the form \cite{KM}
\begin{eqnarray}
\hspace*{-10mm}
A_{l}(t)&=&\!\sum_{j=t,b,\tau}a_{lj}(t)A_{0j}+
\!\!\!\sum_{k=1,2,3}b_{lk}(t)M_{0k}\, ,\qquad l=t,b,\tau
\label{AS} \\
m^2_{n}(t)&=&\!
\sum_{i,j=1,2,3}\!c_{ij(n)}(t)M_{0i}M_{0j}+
\!\!\!\!\!\!\sum_{i,j=t,b,\tau}\!d_{ij(n)}(t)A_{0i}A_{0j}
+\!\!\!\!\!
\sum_{\displaystyle\stackrel{ i=t,b,\tau}{\stackrel{j=1,2,3}{}}}
\!e_{ij(n)}(t)A_{0i}M_{0j}
+\!\sum_{q}k_{q(n)}(t)m^2_{0q}\nonumber
\end{eqnarray}
 where $m^2_{n}$ represent the squark, slepton and
Higgs mass terms,
 $n,q=Q_3,U_3,D_3,H_1,H_2,E_3,L_3$ and $m^2_{0n}$, $A_{0k}$, $M_{0j}$ ,
($k=t,b,\tau ,\,j=1,2,3$) are the initial values of the parameters.
 In one loop order the coefficients of eq.(\ref{AS}) can be
calculated  within the iteration procedure described in \cite{KM}.
We have calculated them up to the 6-th iteration, that allows one
to get accuracy  of 1\%.
 Evaluated at the $M_Z$ scale, which corresponds to
$t=\log M_{GUT}^2/M_Z^2\approx 66$, they depend on the
initial values of Yukawa couplings.
 In what follows we use the
notation $Y_k\equiv h_k^2/16\pi^2 \ (k=t,b,\tau)$ and ${\mbox a}_i
\equiv \alpha_i/4\pi \equiv g_i^2/16\pi^2 \ (i=1,2,3)$.
 To test the behaviour of the coefficients, we take several sets
of Yukawa couplings $Y^0_{t,b,\tau}$ at the GUT scale in the range
$(0.5\div25){\mbox a}_0$ with some arbitrary ratios of
$Y_b/Y_t$ and $Y_\tau/Y_t$  (${\mbox a}_0$ is the common value of
 the gauge couplings at the GUT scale).
  The upper bound for $Y^0_i$ is taken to preserve the
 perturbativity up to GUT scale, the lower one keeps us in
the large $\tan\beta$ regime.

\begin{figure*}[p]
  \hspace*{-.02\textwidth}
\includegraphics[width=.52\textwidth]{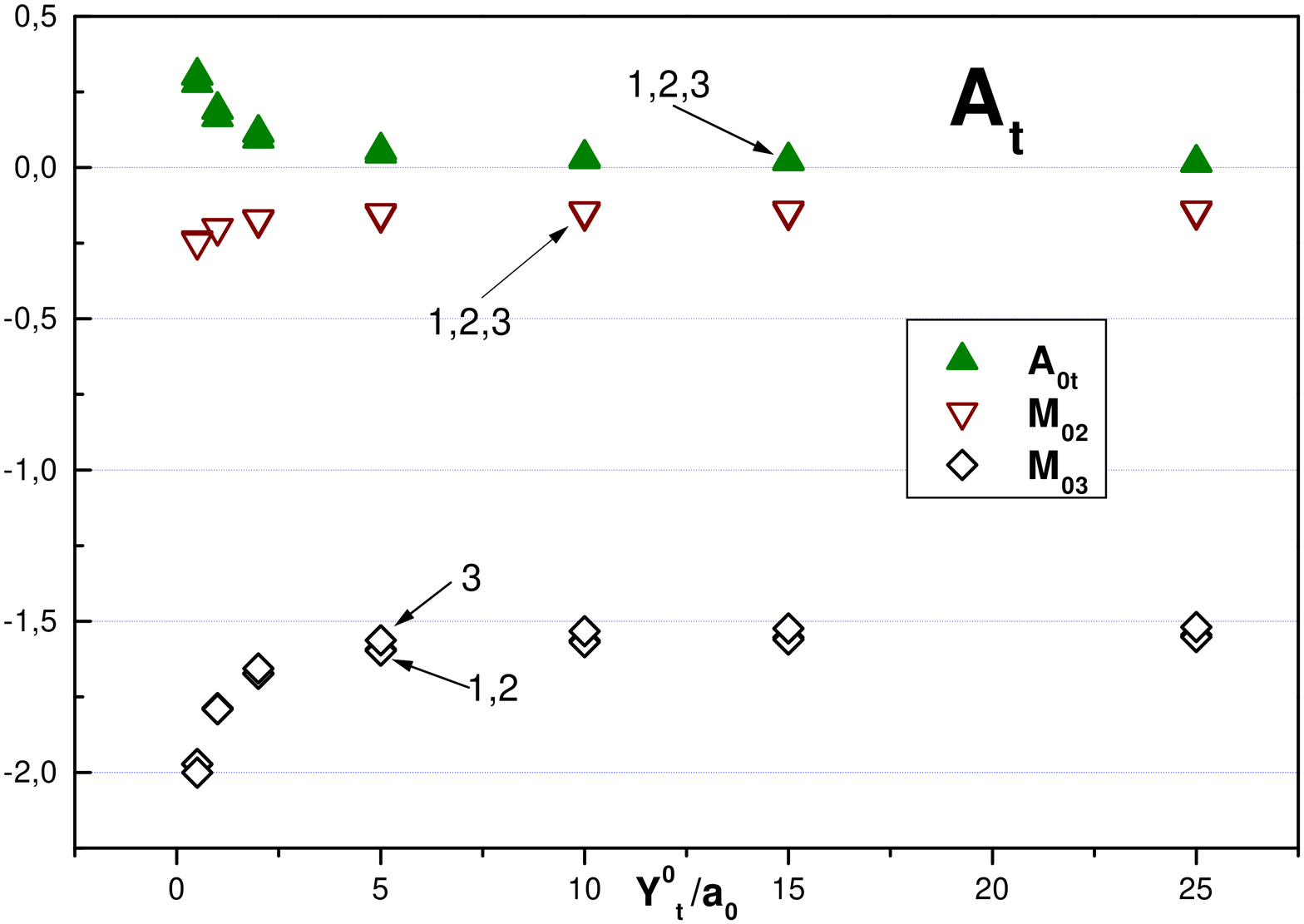}
\hspace*{-.036\textwidth}
\includegraphics[width=.52\textwidth]{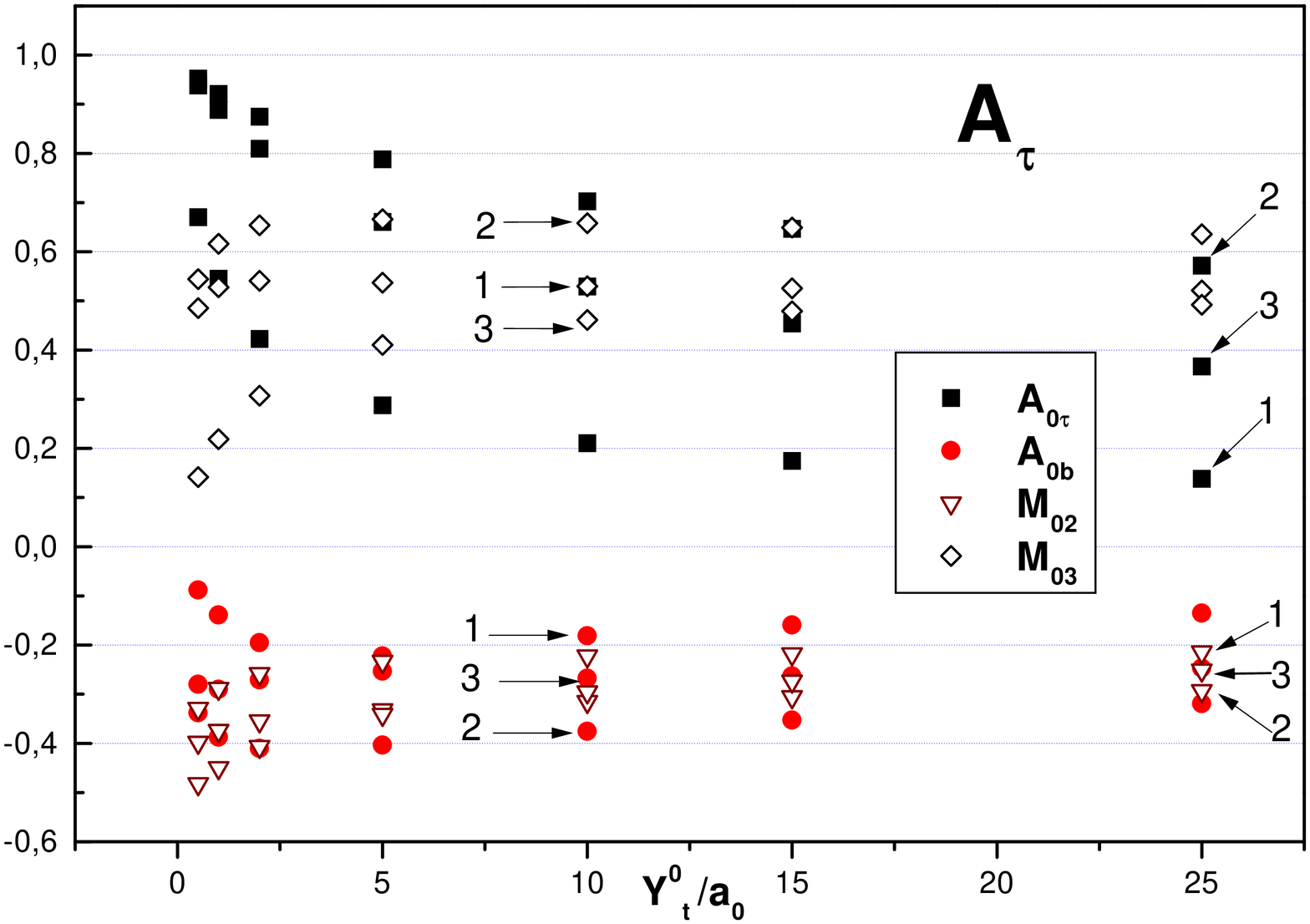}\\
 \hspace*{-.02\textwidth}
\includegraphics[width=.52\textwidth]{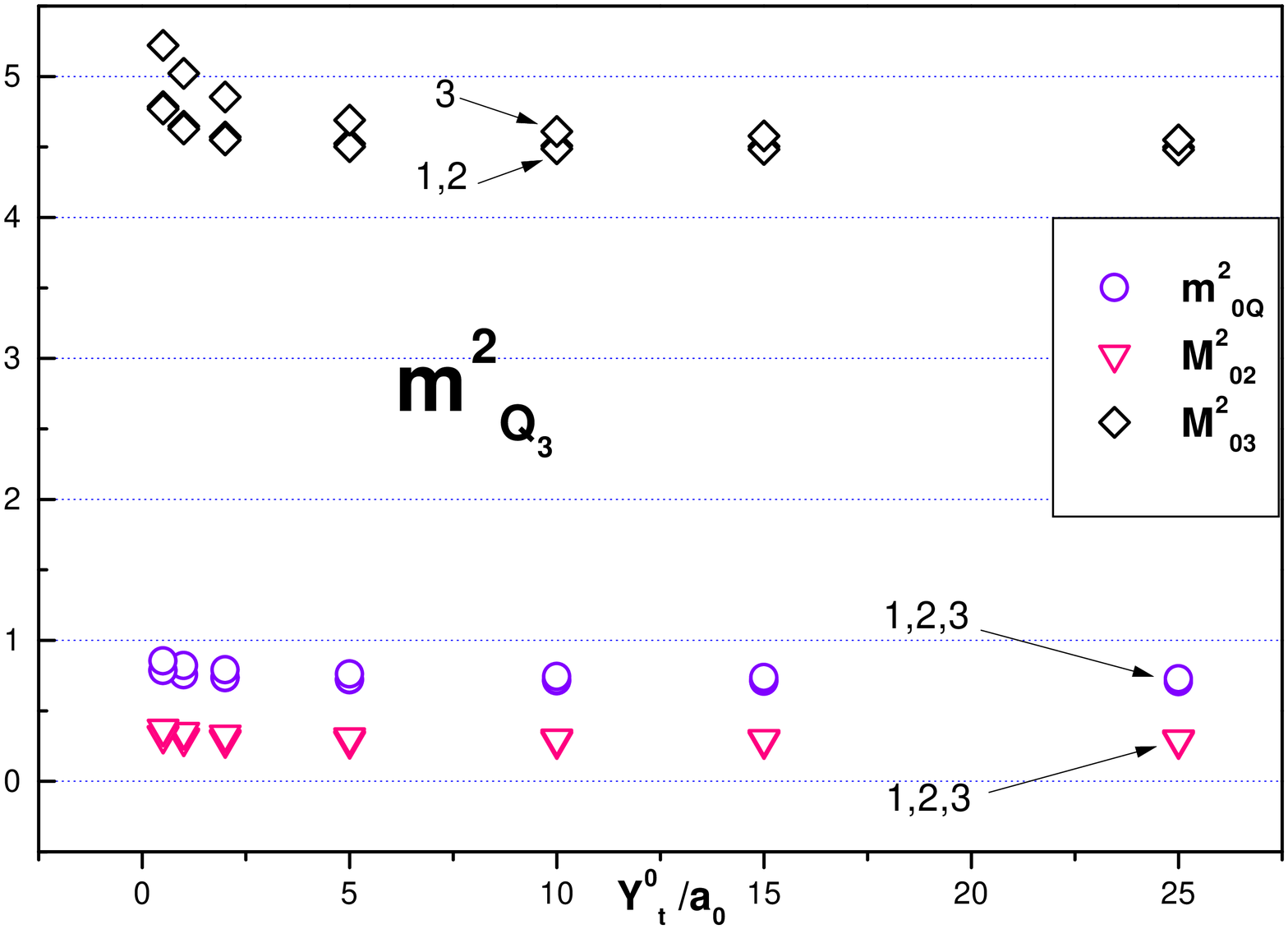}
\hspace*{-.036\textwidth}
\includegraphics[width=.52\textwidth]{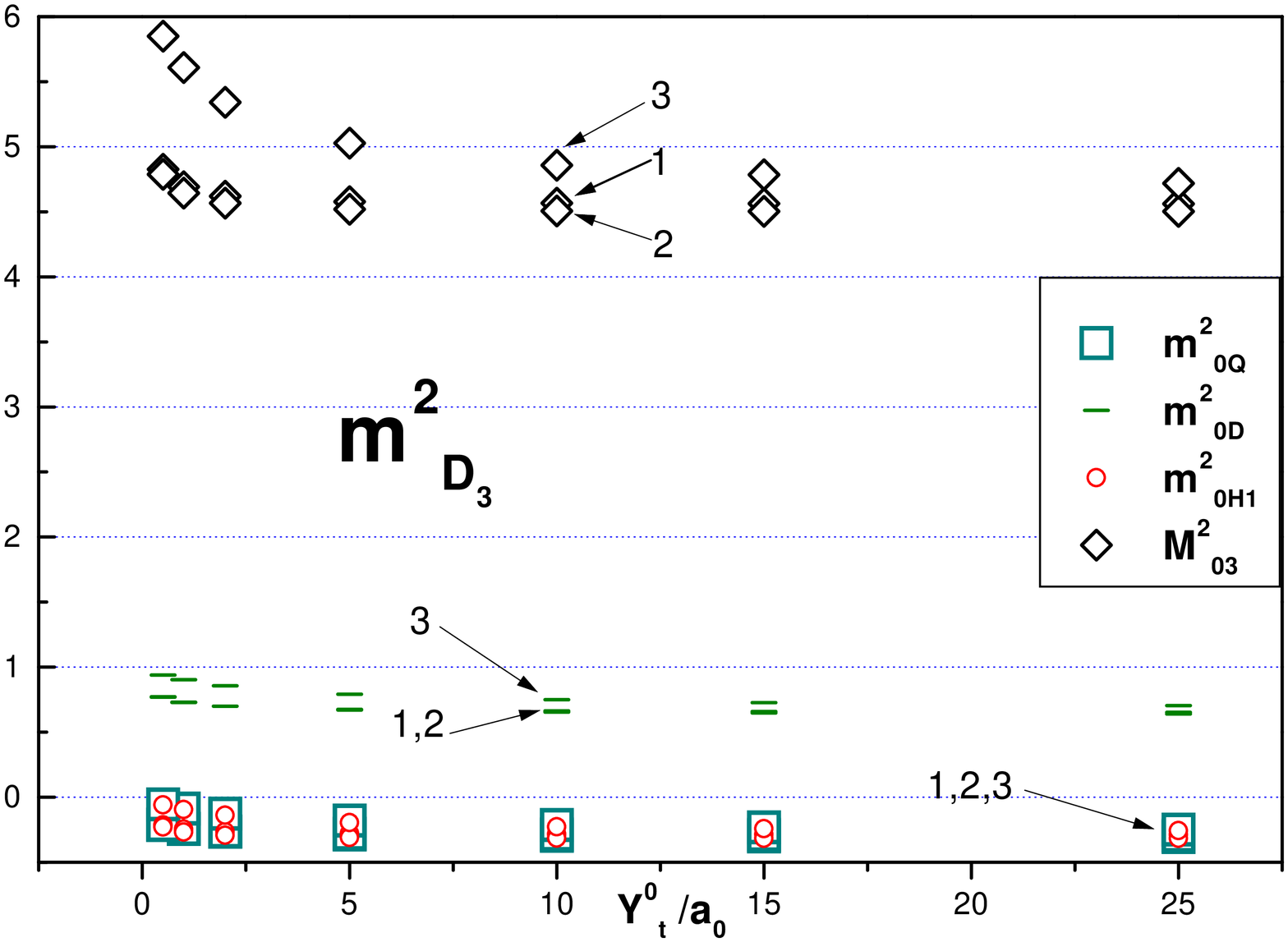} \\
 \hspace*{-.02\textwidth}
\includegraphics[width=.52\textwidth]{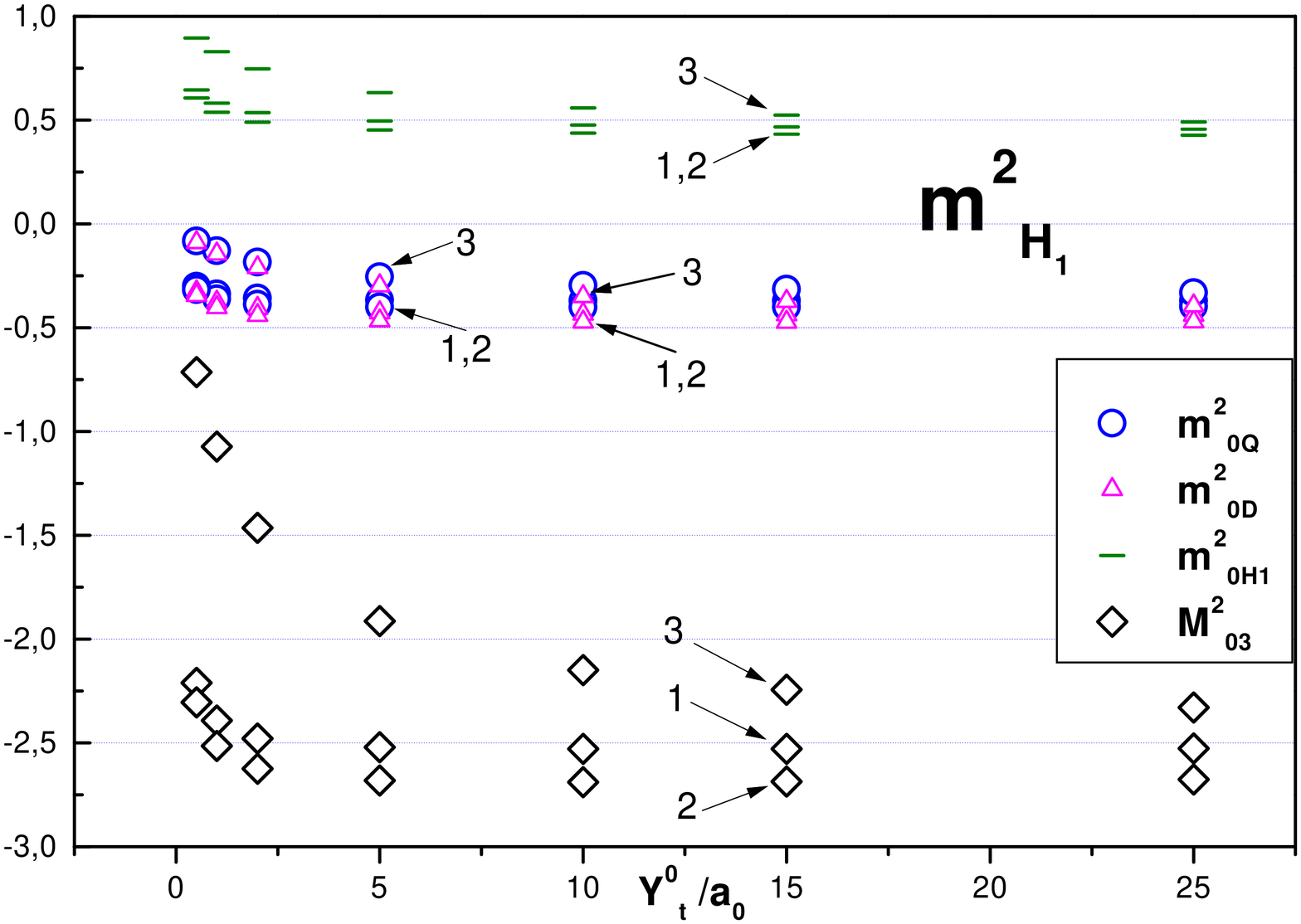}
\hspace*{-.036\textwidth}
\includegraphics[width=.52\textwidth]{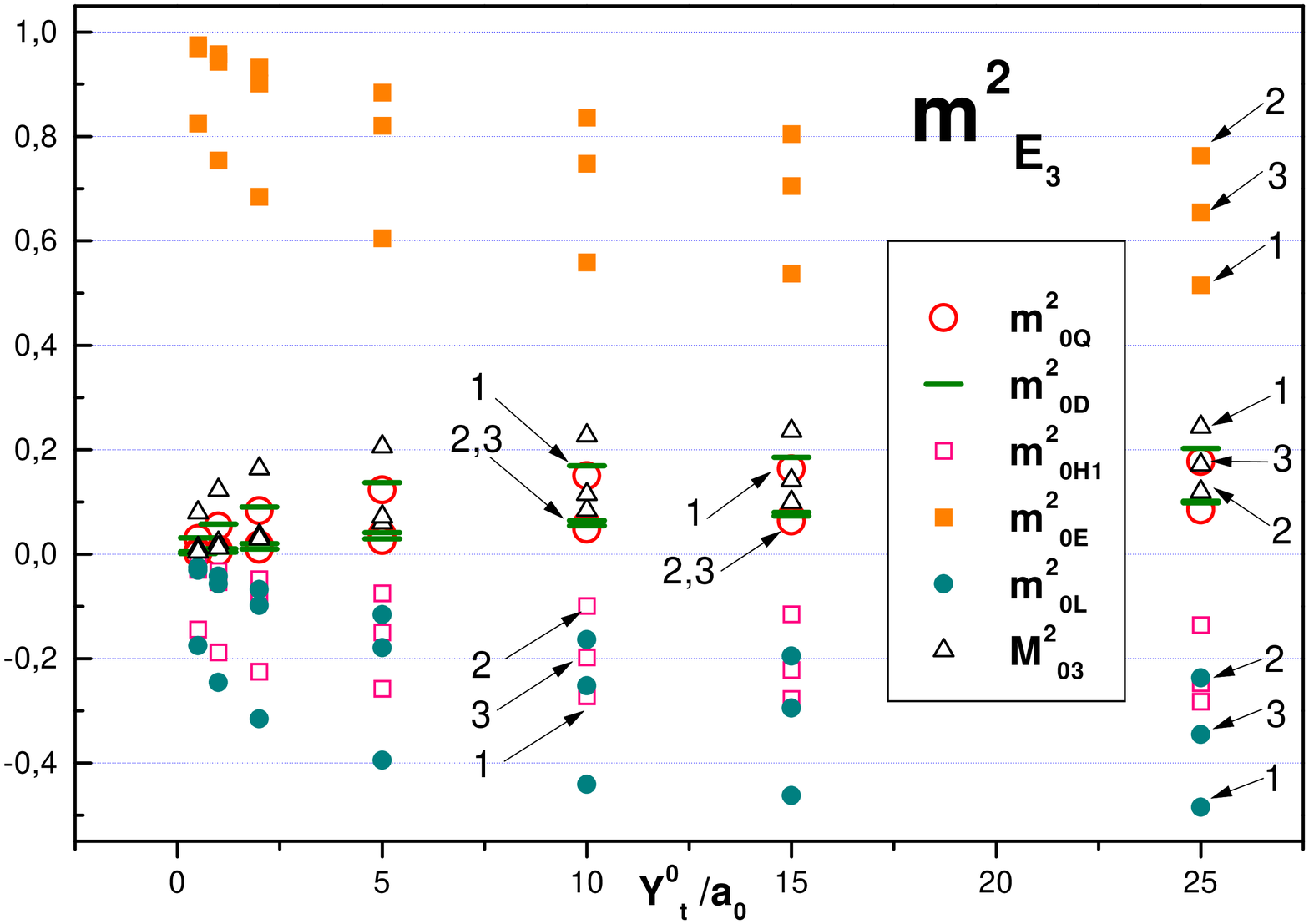}
\caption{
 The dependence of the coefficients of eq.(\ref{AS}) on
the values of Yukawa couplings.
 The coefficients are computed at $t=66$ ($M_Z$ scale).
 For a given $Y^{0}_{t}$ the plotted points
correspond to three different sets of $Y^{0}_{b},Y^{0}_{\tau}$, namely
1): $Y^{0}_{t}=Y^{0}_{b}=Y^{0}_{\tau}$,
2): $Y^{0}_{t}=Y^{0}_{b}=10Y^{0}_{\tau}$,
3): $Y^{0}_{b}=Y^{0}_{\tau}=(1/10)Y^{0}_{t}$.
 Some points for a given  $Y^0_{t}$ may overlap on the plot.
 To keep the plots readable  no connecting lines between points
are drawn, they can be easily recovered due to smooth behaviour.
 For the same reason the coefficients that are close to zero
are not shown. }
\label{a_mssmbig}       
\end{figure*}

 In Fig.\ref{a_mssmbig} we plot the coefficients of eq.(\ref{AS})
as functions of $Y^0_t/{\mbox a}_0$.
 For fixed $Y^0_t$ different  points for a given coefficient
correspond to different relative ratios $Y^0_t/Y^0_{b,\tau}$.
 In this section for illustration we consider three extreme cases:
 $Y^{0}_{t}=Y^{0}_{b}=Y^{0}_{\tau}$,
 $Y^{0}_{t}=Y^{0}_{b}=10Y^{0}_{\tau}$ and
 $Y^{0}_{b}=Y^{0}_{\tau}=(1/10) Y^{0}_{t}$ ,
to demonstrate relative insignificance of the non-universality
of the Yukawa couplings.
Indeed, one can see that for $Y^0_{t}\ge 2{\mbox a_0}$ and regardless of the
ratios $Y^0_t/Y^0_{b,\tau}$ (but still remaining in the range corresponding
to large $\tan\beta$) the coefficients of soft breaking parameters in
$A_{t,b,\tau}$ approach the asymptotic values equal to their IRQFPs.
   Since at the same scale one has
 $M_1(M_Z)=0.412\ M_{01},\,\, M_2(M_Z)=0.822\
 M_{02},\,\,M_3(M_Z)=2.85 \ M_{03}$, we conclude that the ratios
 $A_{t,b}/M_3$ exhibit the proper IRQFP behaviour as used in Ref.\cite{JK}.

 Hence, non-universality of the soft terms changes almost nothing
in the IRQFP behaviour for $A_t$ and $A_b$ because the coefficient
of $M_{03}$ is bigger than the others by a factor of 5 or more.
 One should also note that non-universality of the Yukawa couplings
has a weak effect on the values of the soft  terms coefficients.
On the plot for $A_t$ (the same is true for $A_b$) for a given
coefficient the three points corresponding to various ratios of
 Yukawa couplings almost overlap.
 For $A_\tau$ all the coefficients  but $A^0_{t}$ and $M_{01}$ have
comparable non-vanishing values and the IRQFP is less stringent.
 There is also  stronger dependence on the relative ratios of Yukawa
couplings.
 Fortunately, $A_\tau$ does not play any significant role in the
Higgs mass prediction.

 The same observations holds true for the soft masses.
Taking the values of the Yukawa couplings at the GUT scale as above,
we have found that  for $m^2_{Q_3}$, $m^2_{U_3}$ and $m^2_{D_3}$
the prevalence of the gluino mass $M^2_{03}$ is obvious and
non-universality does not change anything
(see on Fig.\ref{a_mssmbig} the $m^2_{Q_3}$ and $m^2_{D_3}$ plots)
 when comparing with the universality case.

  In the expression of  $m^2_{H_1}$ the coefficients for some  scalar
masses are opposite in sign and of the same magnitude ($m^2_{0H_1}$
comes with '+' sign and $m^2_{0Q_3}$, $m^2_{0D_3}$ come with  '-' sign),
and the same is true for $m^2_{H_2}$ (with $D \to U, H_1 \to H_2$).
  In the case of universal boundary
conditions, for the scalar masses $m^2_{H_1}$ and $m^2_{H_2}$ the only
dependence on the initial conditions left  is that on the gluino mass
 $M^2_{03}$ since the scalar mass $m^2_{0n}$
contributions cancel each other.
 In the case of non-universality with some peculiar choice of
initial conditions some residual dependence on the scalar masses
may appear. Nevertheless, one can still rely on asymptotic
plateau for the coefficients at large Yukawa couplings.

 Again, one observes that for $Y_t^0> 2{\mbox a_0}$
the coefficients approach some asymptotic values and the
dependence on non-universality of Yukawa couplings is rather feeble
for $m^2_{Q_3}$, $m^2_{U_3}$ and $m^2_{D_3}$, and small enough
for $m^2_{H_1}$ and  $m^2_{H_2}$. The residual dependence for
the  coefficients of $M^2_{03}$ in the latter case is because
we get out of the large Yukawa region:
 for $Y^0_t=2{\mbox a_0}$ we have $Y^0_{b,\tau}=0.2{\mbox a_0}$
which doesn't ensure the IRQFP regime.

 The masses of sleptons  $m^2_{E_3}$ and $m^2_{L_3}$ exhibit
rather a fuzzy picture (see the last plot in Fig.\ref{a_mssmbig}).
The coefficient of $M^2_{03}$ is no longer the leading one, instead
we have large contributions from $m^2_{0E_3}$ and $m^2_{0L_3}$.
  Here some coefficients are negative,
thus leading to negative values of e.g. $m^2_{E_3}$ if
$m^2_{0L_3}$ and $m^2_{0H_1}$ are big enough.
 The requirement of positiveness of slepton masses imposes
some bounds on non-universality choice.
 In our analysis below we take the relative ratios of
the soft masses at the GUT scale in the  range $0.5\div 2$
which ensures $m^2_{E_3}>0$ for the most regions
of the parameter space.
 These bounds are in agreement with those
obtained in \cite{bottom-up} in the bottom-up approach.
In the universal case both   $m^2_{E_3}$
and  $m^2_{L_3}$ are positive due to the cancellations between
different soft terms.

On the plots in Fig.\ref{a_mssmbig}, only those parameters which
have non-negligible coefficients are shown.

Thus, we come to the following conclusions:

i) if the Yukawa couplings at the GUT scale are large enough
 ( $> 2 {\mbox a}_0$) the coefficients of eqs.(\ref{AS}) for the soft
terms at low energy scale approach the asymptotic values for both
the universal and non-universal boundary conditions, independently
of the relative ratios of the Yukawa couplings;\\[2mm]
ii) for $A_t,A_b,m^2_{Q3},m^2_{U3},m^2_{D3},m^2_{H1}$ and
$m^2_{H2}$ at the $M_Z$ scale the coefficient of $M_{03}\ (M^2_{03})$
dominates, the IRQFP behaviour is substantiated and can be used
for  both the universal and non-universal boundary conditions.

\section{Low $\tan\beta$ Scenario}\label{lowtan}

We begin our analysis of the influence of non-universality on the
mass of the lightest Higgs boson in the low $\tan\beta$ case.
The present
approach is based on our previous papers \cite{YJK,JK} where we
investigated the mass spectrum of sparticles and the Higgs
bosons using the conception of infrared quasi-fixed points
(IRQFPs) with the
assumption of universality of the soft supersymmetry
breaking parameters. The concept of IRQFPs, which has been
introduced in \cite{Hill}, was widely  employed in the literature
\cite{YJK,JK,WinnyPuh_i_vse_vse_vse,nath_all,HLR}. It allows one
to find the values of the relevant parameters at the $M_Z$ scale
without exact knowledge of their initial values. The validity of
the fixed points is clearly demonstrated in the previous section.
This analysis gives us an important information about the weight
of various initial  parameters in the calculations at low energy
values and, finally, in the calculation of the mass spectrum.

In our previous papers \cite{YJK,JK}, we have concluded that all
Higgs bosons in the MSSM except the lightest CP-even one, are
too heavy to play an important role in the near future experiments;
therefore, in the present paper we concentrate on the
lightest Higgs boson only.

As input parameters we take the known value of the top-quark pole
mass, $m_t^{pole}=173.8 \pm 5.2$ GeV \cite{top}, the experimental
values of the gauge couplings \cite{top} $\alpha_3=0.120 \pm
0.005, \ \alpha_2=0.034,\ \alpha_1=0.017$ and  the sum of the
Higgs vev's squared $v^2 = v_1^2+v_2^2 \approx $174.1 GeV$^2$. We
use the approximate and/or numerical solutions of the relevant RG
equations to evaluate the fixed point values of the mass
parameters. To calculate the mass of the lightest Higgs boson, one
also needs to know the ratio $v_2/v_1$ known as $\tan\beta$ and
the mass parameter $\mu$. To determine $\tan\beta$, we use the
well-known relation between the top-quark running mass, top Yukawa
coupling and $\sin\beta$
\begin{equation}
m_t=h_t v \sin{\beta}. \label{tm}
\end{equation}
The top-quark running mass is found from the pole mass taking into
account QCD and SUSY corrections \cite{mtop1,mtop2} as (for
details see Refs.\cite{YJK,JK})
\begin{equation}
m_t(m_t)=\frac{m_t^{pole}}{1+ \left({\displaystyle\frac{\Delta
m_t}{m_t}}\right)_{QCD} + \left({\displaystyle\frac{\Delta
m_t}{m_t}}\right)_{SUSY}}. \label{pole}
\end{equation}
The results depend on the sign of the $\mu$ parameter which enters the
mixing terms in the stop sector.  For $\mu>0$, one obtains
$m_t(m_t) = 162 \pm 5$ GeV. Negative values of $\mu$ lead to a too
light Higgs boson, and we do not consider this case here.

Now we can estimate the value of the $\tan\beta$. We assume  that
the top Yukawa coupling is close to its IRQFP. This is realized
when $ \rho_{t0}=Y_t^0/{\mbox a}_0 >2$. Then one gets
$h_t(M_Z)=1.09-1.14$ when $2< \rho_{t0} <25$ .
As a central value of top Yukawa coupling we take
$h_t(M_Z)=1.12$ which corresponds
to $\rho_{t0}=5$. This gives the following value of $\tan\beta$:
$$\tan\beta = 1.47 \pm 0.1 \pm 0.15 \pm 0.05, \ \ \ \ \mu>0 . $$
The deviations from the central value are connected with the
deviation from the fixed point value of the Yukawa coupling and
the  experimental uncertainties in the top-quark mass and
$\alpha_3(M_Z)$, respectively.

The Higgs mixing parameter $\mu$ can be determined from the
requirement of radiative EWSB  and can be found from the Higgs
potential minimization condition. In contrast with our previous
paper \cite{YJK} (where we have taken into account only tree level
minimization condition), we include here the one-loop corrections.
It gives the difference of about 2 GeV for the lightest Higgs
boson mass. The one-loop minimization condition  reads
\begin{equation}
\frac{M_Z^2}{2}+\mu^2=\frac{m_{H_1}^2+\Sigma_1-
(m_{H_2}^2+\Sigma_2) \tan^2\beta}{\tan^2\beta-1}\,,
\label{MZC}
\end{equation}
where $\Sigma_1$ and $\Sigma_2$ are the one-loop corrections
\cite{Gl}, $M_Z$ is the $Z$-boson mass and $m_{H_1}^2$ and
$m_{H_2}^2$ are the Higgs soft mass parameters which are
determined by solutions of the RG equations. The latter possess
the IRQFP's which we use in our analysis. The above equation
allows one to obtain an absolute value of $\mu$. The sign of
$\mu$ remains a free parameter, however, as it has already been
mentioned, negative values of $\mu$ give too small values of the
lightest Higgs boson mass and are excluded experimentally
\cite{LEP_new_results}.

In the MSSM, the Higgs sector consists of five physical states: two
neutral CP-even scalars $h$ and $H$, one neutral CP-odd scalar
$A$, and two charged Higgs scalars $H^{\pm}$. We concentrate on
the  mass of the lightest Higgs boson $h$. At the tree level, the
mass of $h$ is smaller than the mass of $Z$-boson, $M_Z$, but the
loop corrections increase it. In general, the mass matrix for the
CP-even neutral  Higgs bosons looks like
\begin{eqnarray}
{\mathcal M}\!\!&=&\!\!\left(\!\!\begin{array}{cc}\tan\beta & -1\\ -1 &
\cot\beta
\end{array}\!\! \right)m^2_A\cos\beta\sin\beta
+  \left(\!\!\begin{array}{cc}\cot\beta & -1\\ -1 & \tan\beta
\end{array}\!\!\right)M^2_Z\cos\beta\sin\beta
 +  \left(\!\!\begin{array}{cc} \Delta_{11} & \Delta_{12}\\
\Delta_{12} & \Delta_{22}
\end{array}\!\!\right)  \label{h}
\end{eqnarray}
where $m_A$ is the mass of the CP-odd Higgs boson and $\Delta's$
are the radiative corrections~\cite{cor}. These corrections depend
on stop masses which are given by the following equation:
\begin{eqnarray}
\tilde m_{t_{1,2}}^{2}&=&\frac{1}{2} \big[\tilde m_{t_L}^{2}
+\tilde m_{t_R}^{2} \mp \sqrt{(\tilde m_{t_L}^{2} - \tilde
m_{t_R}^{2})^2 +4 m_{t}^{2} (A_t-\mu \, \cot \beta)^2} \big] \,,
\label{stop}
\end{eqnarray}
\begin{figure}[t]
\includegraphics[width=.5\textwidth]{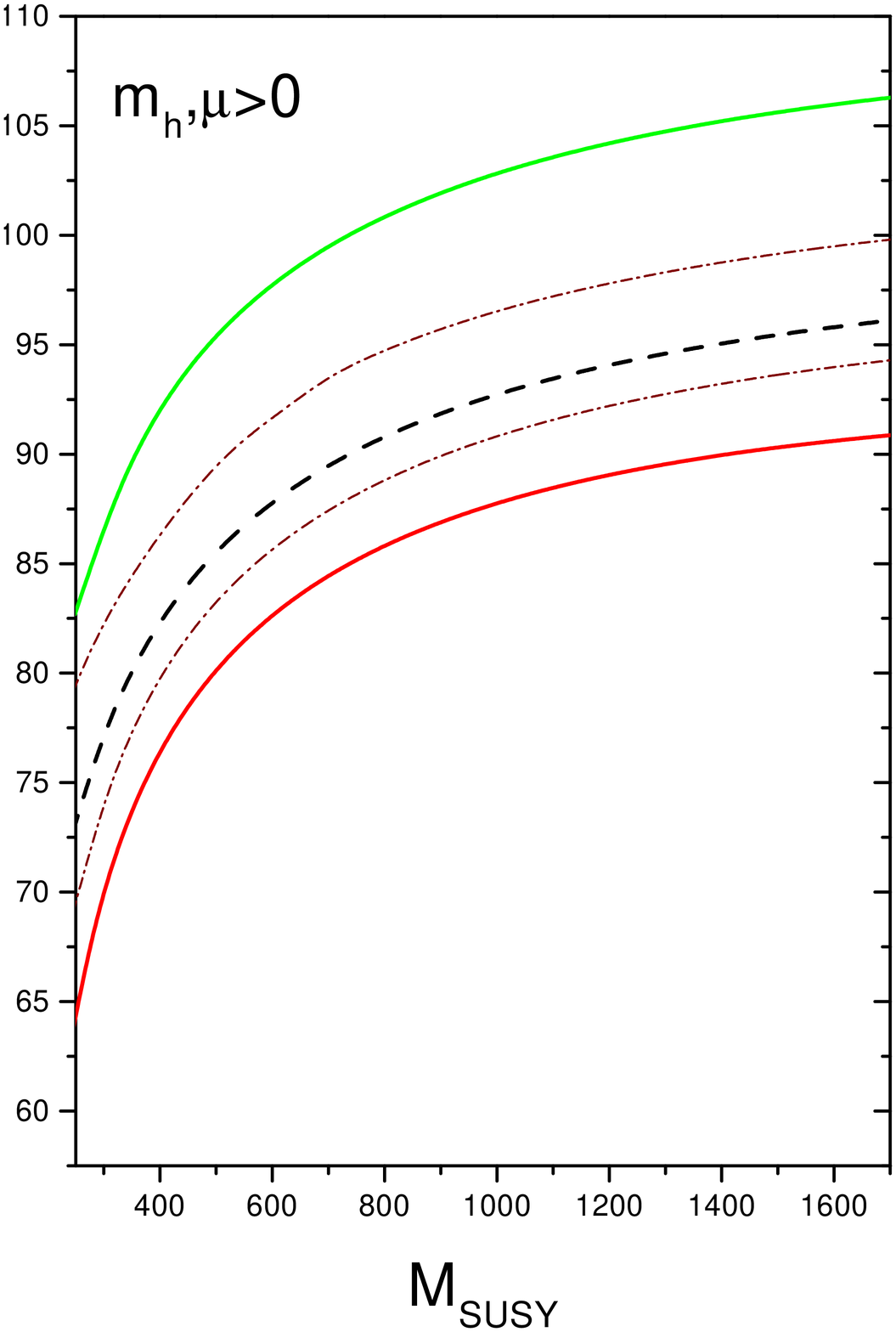}
\hspace*{-.015\textwidth}
\includegraphics[width=.5\textwidth]{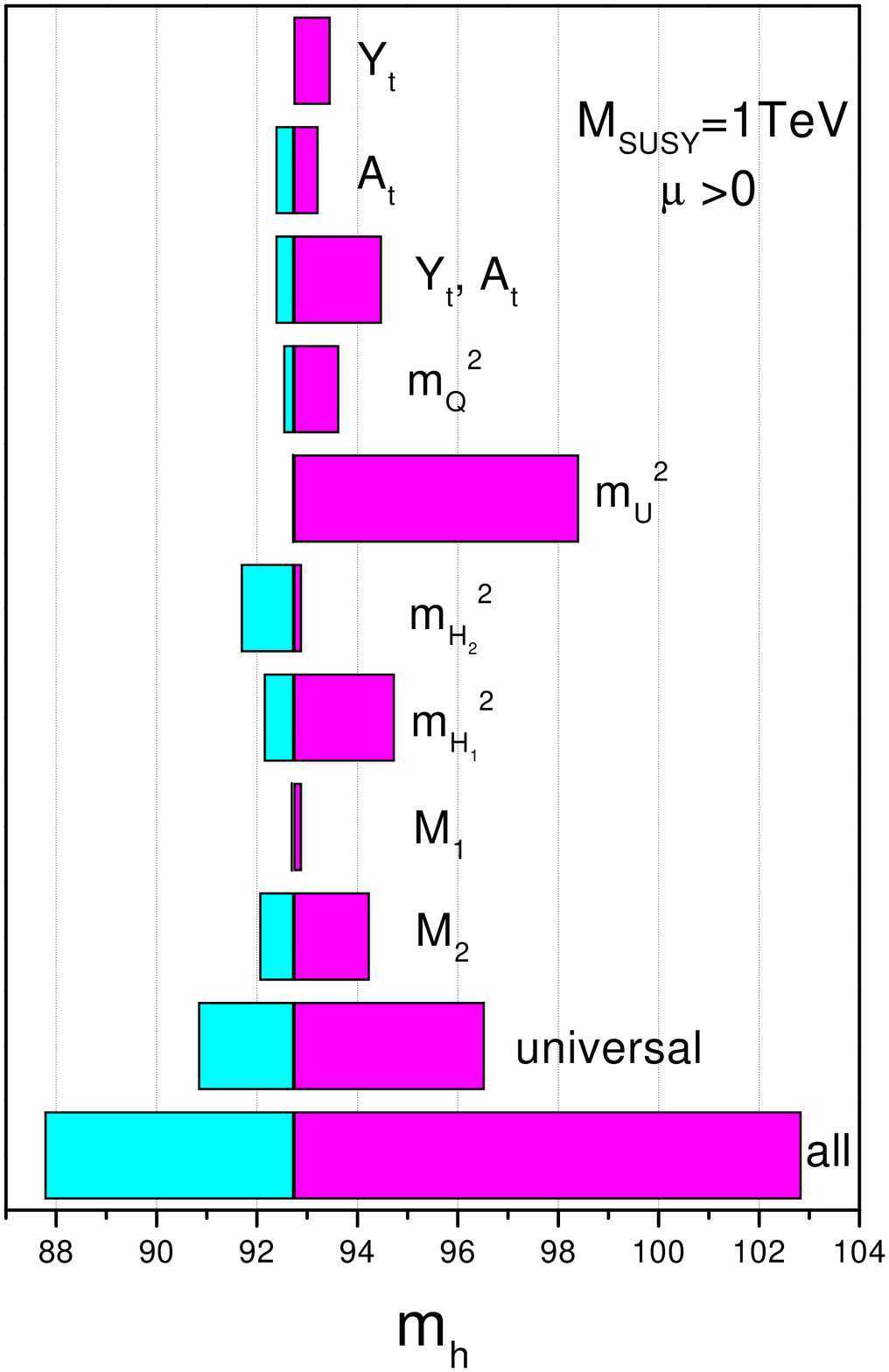} \\
\vspace*{-10mm}
\caption{  The mass of the lightest Higgs boson as a function of
$M_{SUSY}$. The dashed (central) line  corresponds to the central
values of the parameters, the dash-doted lines corresponds to the upper
and lower limits in the case of universal boundary conditions and
the solid lines are absolute upper and lower limits on the mass of
the lightest Higgs boson in the non-universal case (left).
 The influence of variations  from  central
values of the individual parameters as well as their collective
effect on the mass of the lightest Higgs boson  in both the
universal and non-universal cases at a typical scale $M_{SUSY}=1$
TeV (right). \label{fm}}
  \end{figure}
To find the Higgs boson mass, one has to diagonalize the mass
matrix (\ref{h}). In our previous paper \cite{YJK} we have
estimated the mass of the lightest Higgs boson exploring the
IRQFPs in case of universal boundary conditions. In the present
paper, we also use the concept of IRQFPs but  release the
universality conditions and allow moderate deviations from
universality. In view of the analysis of the previous section, we
expect that the main influence of non-universality comes from the
initial values of the Higgs masses $m^2_{H_1}$ and $m^2_{H_2}$ and
that of $m^2_{U}$ while those of $m^2_{Q}$ and gauginos are of
minor importance. In the numerical analysis,
we consider the following intervals for the top Yukawa coupling
and soft breaking parameters at the GUT scale:
$Y_{t}^0/{\mbox a}_0 \in <2,25>$, $A_{0t}/M_{03} \in
<-1,1>$, $m_{0i}^2/M_{03}^2 \in <0.25,4>$ and $M_{0j}/M_{03} \in
<0.5,2>$, where  $i=(Q_3, U_3, H_1, H_2)$ and $j=1,2$.

In Fig.\,\ref{fm}, the dependence of the mass of the
lightest Higgs boson $m_h$ on the geometric mean of stop masses
$\sqrt{\tilde m_{t_1} \tilde m_{t_2}}$ is shown,
 which is often identified
with the supersymmetry breaking scale $M_{SUSY}$. One can see
obvious saturation of the Higgs mass when $M_{SUSY} \geq 500$
GeV. The central (dash) line corresponds to the central values of
the parameters. We take them as follows: $Y_{t}^0/{\mbox a}_0 =5$,
$A_{0t}/M_{03}=0$, $m_{0}^2/M_{03}^2=1$ for all scalar masses  and
$M_{0j}/M_{03}=1$ for gaugino masses  $M_{01}$ and $M_{02}$.  If
one assumes the universality, the mass of the lightest Higgs boson
at a typical scale $M_{SUSY}=1$ TeV  ($\mu>0$) is
\begin{equation}
m_h=92.7 \ ^{\displaystyle +3.8}_{\displaystyle -1.9}\  \pm5\
\pm0.4 \ \mbox{GeV}, \ \ \ \mbox{ for} \ M_{SUSY}=1 \ TeV.
\label{mass}
\end{equation}
The first uncertainty is given by the deviations from central
values of the top Yukawa coupling and soft breaking parameters
($+3.8\ (-1.9)$), the second one by the
uncertainty of top-quark mass
and the third one by uncertainty in the strong coupling constant
$\alpha_3(M_Z)=0.120 \pm 0.005$.
If the parameters are non-universal at the GUT scale, the range of
possible values of the lightest Higgs boson mass becomes wider:
\begin{equation}
m_h=92.7 \ ^{\displaystyle +10.1}_{\displaystyle -4.9}\ \pm5 \
\pm0.4 \ \mbox{GeV}, \ \ \ \mbox{ for} \ M_{SUSY}=1 \ TeV.
\label{mass2}
\end{equation}
{}From  Fig.\,\ref{fm}, one can see that the main deviations from
the universal case for the lightest Higgs boson mass is due to the
soft mass parameters, especially $m_U^2$.
One can  see that the
restrictions on the lightest Higgs boson mass in the non-universal
case are not so strict as in the universal one.
 In the non-universal
case, in the MSSM with low $\tan\beta$  it is possible that the
lightest Higgs boson has the mass slightly higher than $100$ GeV
contrary to the universal case.
 However, it is still too light in
view of recent experimental data, which sets the lower limit on
the Higgs mass of 103 GeV \cite{LEP_new_results}.

  Of course, if one allows the parameters to have larger deviations than
we use in present analysis
(we have imposed the following restrictions on the soft masses:
${\displaystyle m^2_{0i}/M^2_{03} \in <.25, 4> }$ and
${\displaystyle m^2_{0i}/m^2_{0j}}$ $ \in <1/16, 16> $ where
$i,j=Q_3,U_3,D_3,H_1,H_2$) it is possible to find the mass
 of the lightest Higgs boson even larger than our upper bound.
 For instance if one allows the soft  masses to be in interval
${\displaystyle m^2_{0i}/m^2_{0j}\in <1/100, 100> }$
than the upper bound for $m_h$ increases by about 3 GeV.
 However, we consider such a large non-universality to be unnatural.
\begin{figure}[t]
\includegraphics[width=.5\textwidth]{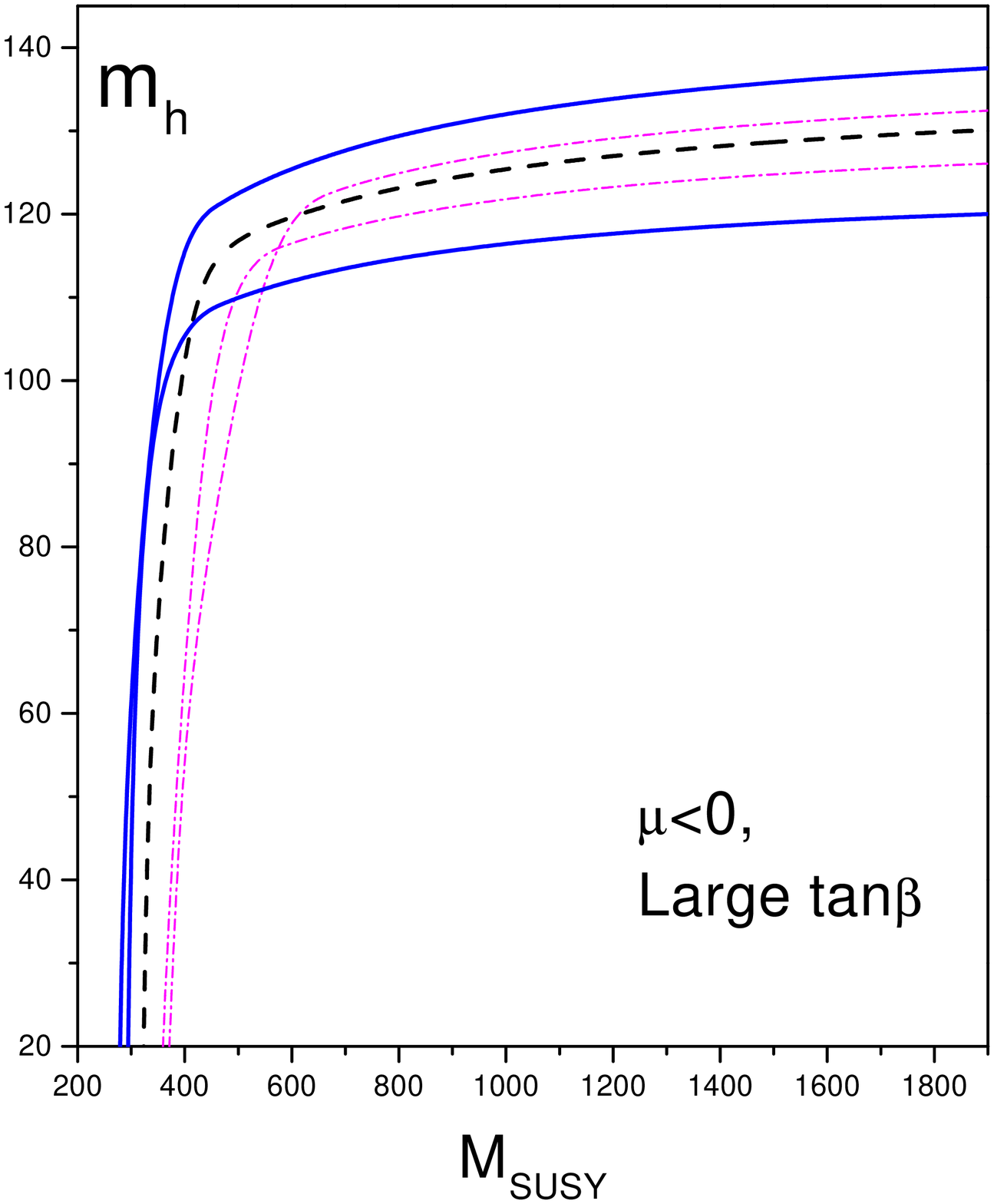}
\hspace*{-.015\textwidth}
\includegraphics[width=.5\textwidth]{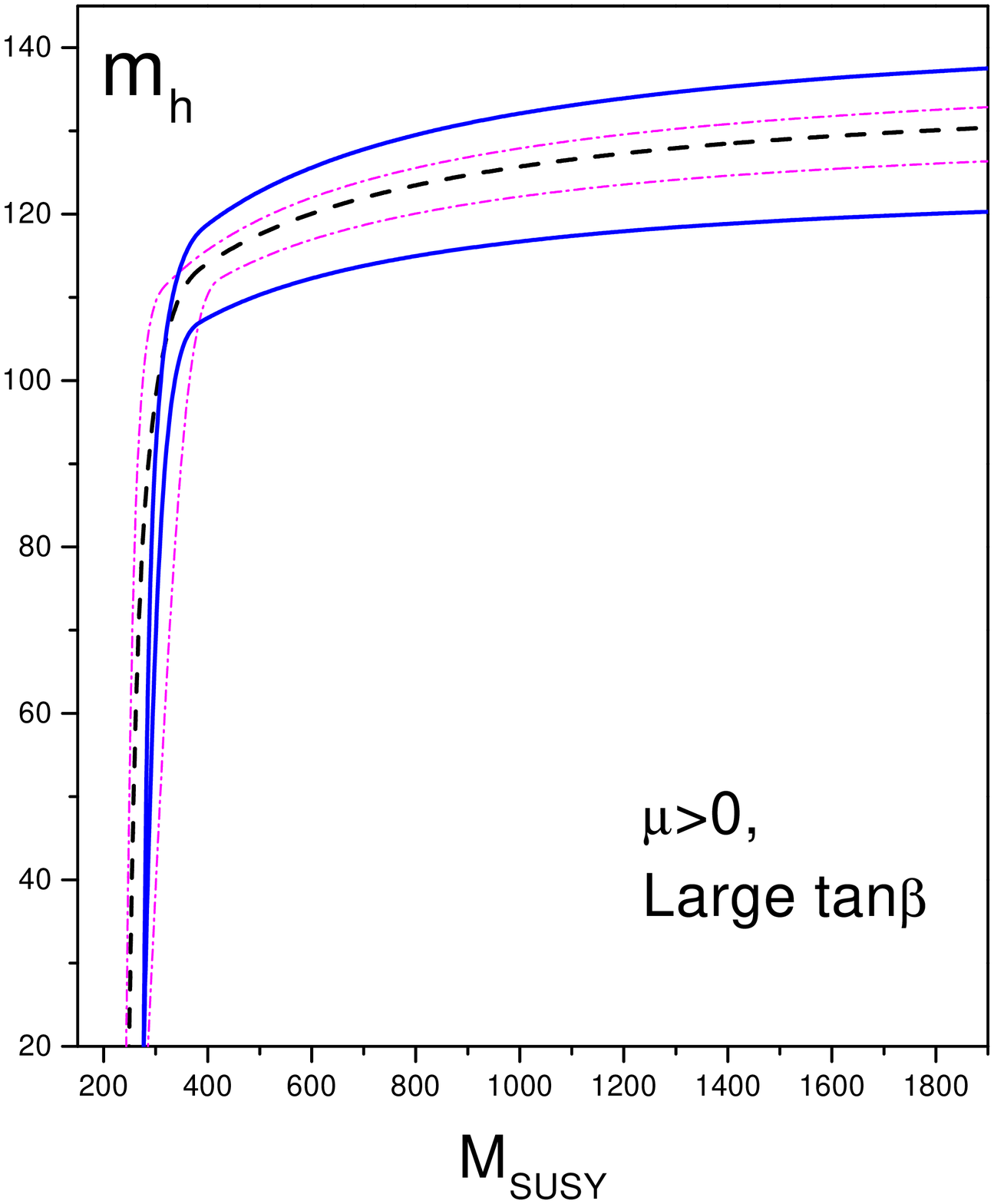} \\
\vspace*{-10mm}
\caption{ The mass of the lightest Higgs boson as
a function of $M_{SUSY}$ for both the cases
$\mu>0$ and $\mu<0$.
 The dashed (central) line corresponds to the central values of the
parameters, the dash-doted lines correspond to the upper and lower
limits for the universal boundary conditions and the solid lines
are the absolute upper and lower limits on the mass of the
lightest Higgs boson in the non-universal case.
The line intersection is related to a steep fall of $m^2_h$
at low values of $M_{SUSY}$ where  $m^2_h$ becomes negative.
The position of the ``switch'' depends on the choice of
parameters.
Physically relevant parts of the plots start at
approximately $M_{SUSY}\ge 400$ GeV ($\mu >0$) and
$M_{SUSY}\ge 600$ GeV ($\mu <0$) \label{fm1} }
  \end{figure}
\section{Large $\tan\beta$ scenario}\label{bigtan}
Consider now  the large $\tan\beta$ case. The  situation is more
complicated because the space of input parameters is larger. We
follow the  same strategy as in Ref.\,\cite{JK}, but with some
modifications. To take into account the non-universality, we keep
the initial values of top and bottom Yukawa couplings within the
whole interval $\rho_{t0},\rho_{b0} \in <2, 25>$ (see the previous
section for notation). In Ref.\,\cite{JK}, we restricted this
interval imposing some constraints on the $\sin\beta$ and
$\tau$-lepton mass $m_{\tau}$. Here, the only restriction is the
attraction to the IRQFPs which defines the above mentioned
interval. To determine $\tan\beta$, we use  equation (\ref{tm}) and
a similar one for the bottom-quark running mass
\begin{equation}
m_b=h_b v \cos{\beta}, \label{tb}
\end{equation}
so that  $\tan\beta$ is defined from
\begin{equation}
\tan\beta=\frac{m_t}{m_b}\frac{h_b}{h_t}. \label{tan}
\end{equation}

The  top-quark running mass has been calculated in
the previous section. As for the bottom-quark running mass,
the
situation is more complicated because the mass of the bottom $m_b$
is essentially smaller than the scale $M_Z$; so we have to take
into account the running of this mass from $m_b$ to the $M_Z$ scale.
The bottom-quark pole mass is $m_b^{pole}=4.94\pm0.15$ GeV
\cite{dev}.
 To calculate the running mass, we use the  well-known
procedure given e.g. in Refs.\,\cite{JK,mtop2,ar,gr}.
Since we
assume non-universality of the Yukawa couplings, the value of
$\tan\beta$ obtained from eq.(\ref{tan}) belongs to a wider
interval.
For the central values of the Yukawa couplings and the
mass parameters (see the previous section) we find the following
values of $\tan\beta$: $\tan\beta=69.3, \mu>0$ and
$\tan\beta=38.1, \mu<0$.
 When the parameters vary around their
central values, $\tan\beta$ is changing within the  intervals:
$$\tan\beta \in <41.2, 130.2> \ \  (\mu>0), \  \quad  \ \
\  \tan\beta \in <35.0, 40.6>\ \ (\mu<0).$$
The parameter $\mu$ is
calculated in the same manner as in the low $\tan\beta$ case.
The same is true for the stop and sbottom masses.
\begin{figure}[t]
\includegraphics[width=.5\textwidth]{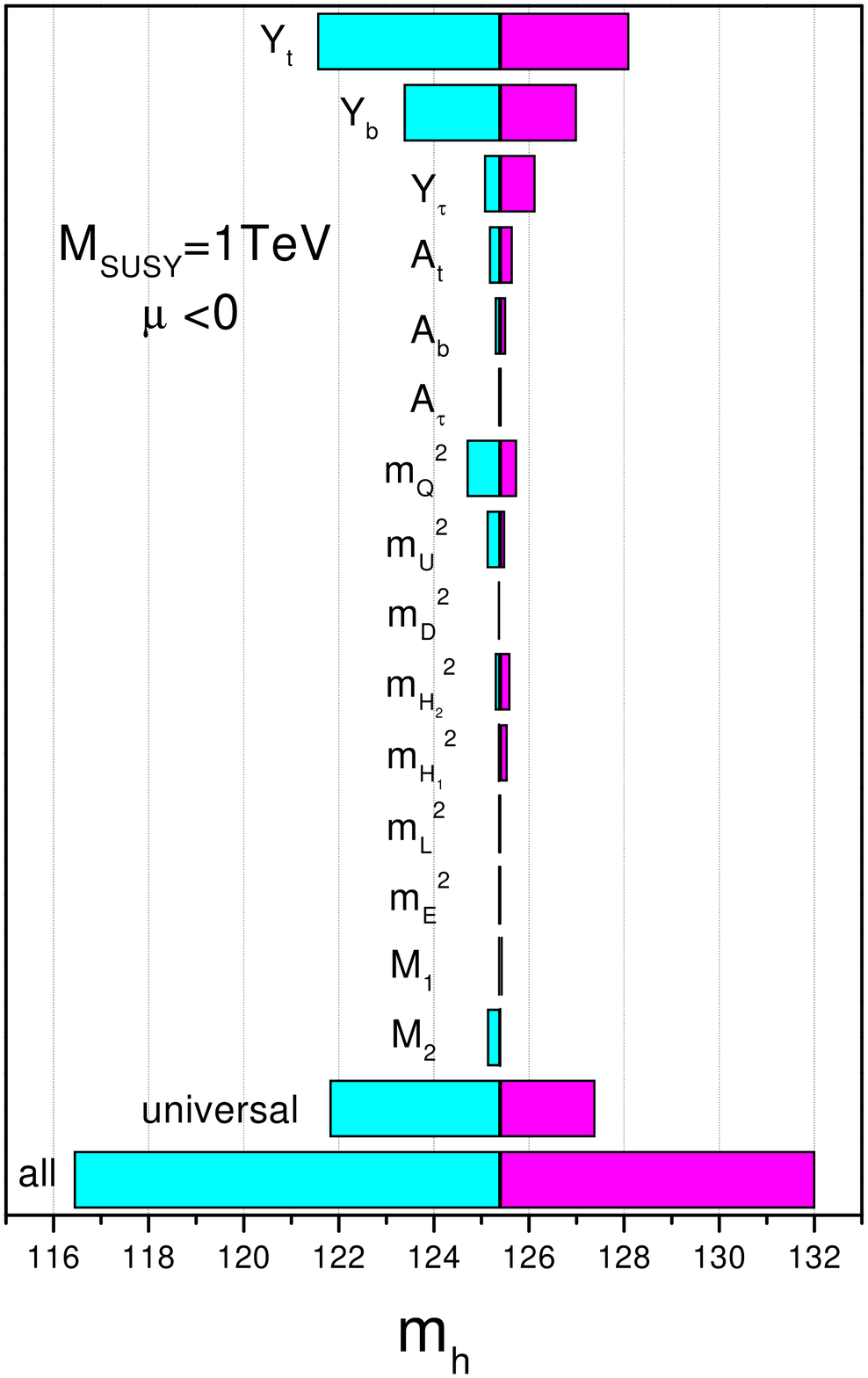}
\hspace*{-0.015\textwidth}
\includegraphics[width=.5\textwidth]{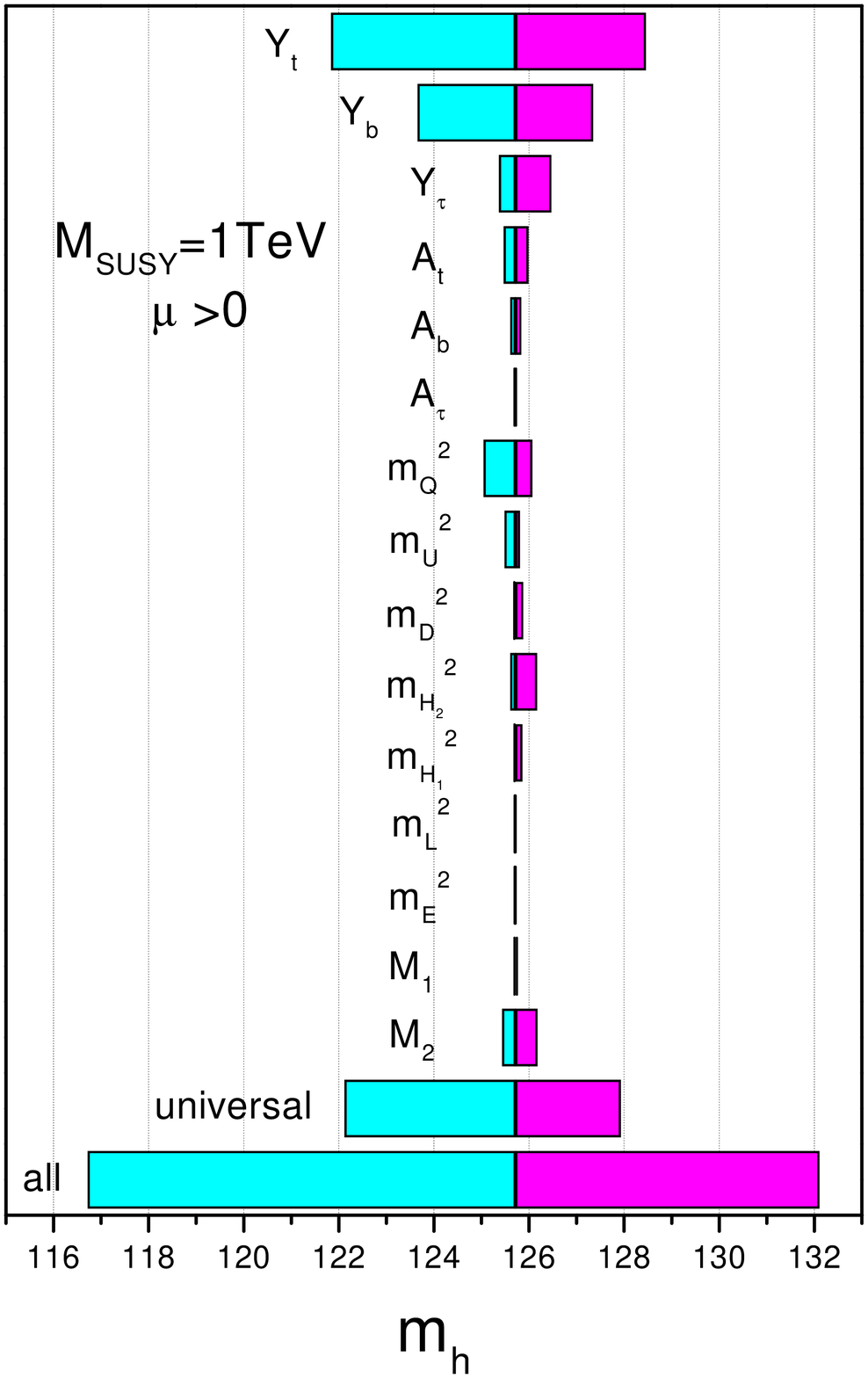}
\caption{ The influence of the variations from  central values of
the  individual parameters and their collective effect on the mass
of the lightest Higgs boson in both the universal and
non-universal cases at a typical scale $M_{SUSY}=1$ TeV.
\label{fm2}
 }
  \end{figure}
In Fig.\,\ref{fm1}, we present the dependence of the mass of the
lightest Higgs boson on $M_{SUSY}$ for both the cases $\mu>0$ and
$\mu<0$.
One can immediately see the very sharp increase of $m_h$
to the plateau starting from $M_{SUSY} \geq 500$ GeV.
 In the non-universal case,
 the interval of masses is slightly wider than
in the universal one.
Fig.\,\ref{fm2} shows  the dependence of the lightest Higgs
boson mass on the deviations of
the individual parameters from their central values for both the
cases $\mu>0$ and $\mu<0$.
 The major influence on the Higgs mass
is given by the Yukawa top and bottom couplings.
 The influence of the
other parameters is negligible.
 One can immediately see the big difference between the universal
 and non-universal cases.
 If one assumes the universality of the Yukawa couplings and soft
parameters,  the mass of the lightest Higgs boson at the typical
scale $M_{SUSY}=1$ TeV is given by
\begin{eqnarray}
m_h&=& 125.7 \ ^{\displaystyle +2.2}_{\displaystyle -3.5}\ \pm5 \
\pm0.4 \ \ \  \mbox{GeV for} \ \ \  \mu>0 \,, \nonumber \\
m_h&=&125.4 \ ^{\displaystyle +2.0}_{\displaystyle -3.6}\ \pm5 \
\pm0.4 \ \ \ \mbox{GeV for} \ \ \ \mu<0 \,. \nonumber
\end{eqnarray}
The first uncertainty is connected with the  deviations of the
Yukawa couplings and soft parameters from their central values in
the universal case, the second one is due to the experimental
uncertainty in the top-quark mass, and the third one is connected
to that of the strong coupling constant.
 When one does not assume universality,  the allowed interval of
 the Higgs boson mass is wider. For $M_{SUSY}=1$ TeV
we get
\begin{eqnarray}
m_h&=& 125.7 \ ^{\displaystyle +6.4}_{\displaystyle -9.0}\ \pm5 \
\pm0.4 \ \ \ \mbox{GeV for} \ \ \  \mu>0 \,, \nonumber \\
m_h&=&125.4 \ ^{\displaystyle +6.6}_{\displaystyle -9.0}\ \pm5 \
\pm0.4 \ \ \ \mbox{GeV for} \ \ \ \mu<0 \,. \nonumber
\end{eqnarray}
 One can see that in the case of large $\tan\beta$  the mass of the
lightest Higgs boson  typically belongs to the interval $<115,
130>$ GeV. The upper bound on $m_h$ is reached for $Y^0_t$
close to its perturbative limit ($Y^0_t/{\mbox a_0}\approx 25$).
 The influence of the soft parameters is small as one can see in
Fig. \ref{fm2} and is also restricted by the assumption
that the soft masses for sleptons $m^2_{E3}$ and $m^2_{L3}$
are positive at $M_Z$ scale.
  The lower bound on $m_h$ decreases by about 3 GeV when
 the assumption of IRQFP for Yukawa couplings is released.
There is still a constraint on $Y^0_t$ ($Y^0_t/{\mbox a_0} > 1.2$)
 given by condition
$\sin\beta\le 1$ in the relation for top mass (\ref{tm}).
Experiments are still far away from these values,
though the lower boundary may be within the reach of LEP II.

\section{Conclusion}\label{the_end}

We have analyzed the influence of  non-universality of the Yukawa
couplings and soft SUSY breaking parameters on the mass of the
lightest Higgs boson $h$ in the MSSM. Possible values of the Higgs
mass are obtained. This may be important for the Higgs searches in
the nearest future.

In the low $\tan\beta$ case, the main role is played by
non-universality of the mass soft parameters. Assuming a moderate
deviation from universality one gets the mass of the lightest
Higgs boson below 103 GeV which is almost excluded by recent
experimental data \cite{LEP_new_results}.

For high $\tan\beta$ the situation is different.
 Here, the main role is played by non-universality of the
Yukawa couplings;
the variations of the soft terms are of minor importance.
The mass of the lightest Higgs boson in this case is much larger.
Here more interesting
 is the lower bound of the Higgs mass. The effect of
non-universality is the decrease in this bound which may become as
low as 115 GeV leaving hopes for the imminent observation of the
Higgs boson.

\vglue 0.5cm
 {\bf Acknowledgments}

\vglue 0.5cm

We are  grateful to G. Moultaka for useful discussions. Financial
support from RFBR grants \# 99-02-16650 and \# 96-15-96030   is
kindly acknowledged.


\end{document}